\documentclass[
 aps, pra,
 amsmath,amssymb,
 11pt,
 final,
tightenlines,
 twoside,
 twocolumn,
 nofloats,
nofootinbib,
 superscriptaddress,
showkeys,
showkeywords,
 ]
{revtex4-2}

\usepackage[T1]{fontenc}
\usepackage[utf8x]{inputenc}
\usepackage[english]{babel}
\usepackage{graphicx}% Include figure files
\usepackage{dcolumn}% Align table columns on decimal point
\usepackage{bm}% bold math

\input{maik.rty}

\setcitestyle{authoryear,round}
\def\saoname{Special Astrophysical Observatory,  Russian Academy of Sciences,
              Nizhnii Arkhyz, 369167 Russia}

\def\squareforqed{\hbox{\rlap{$\sqcap$}$\sqcup$}}

\def\sq{\ifmmode\squareforqed\else{\unskip\nobreak\hfil
\penalty50\hskip1em\null\nobreak\hfil\squareforqed
\parfillskip=0pt\finalhyphendemerits=0\endgraf}\fi}

\def\utw{\smash{\rlap{\lower5pt\hbox{$\sim$}}}}

\def\udtw{\smash{\rlap{\lower6pt\hbox{$\approx$}}}}

\def\diameter{{\ifmmode\mathchoice
{\ooalign{\hfil\hbox{$\displaystyle/$}\hfil\crcr
{\hbox{$\displaystyle\mathchar"20D$}}}}
{\ooalign{\hfil\hbox{$\textstyle/$}\hfil\crcr
{\hbox{$\textstyle\mathchar"20D$}}}}
{\ooalign{\hfil\hbox{$\scriptstyle/$}\hfil\crcr
{\hbox{$\scriptstyle\mathchar"20D$}}}}
{\ooalign{\hfil\hbox{$\scriptscriptstyle/$}\hfil\crcr
{\hbox{$\scriptscriptstyle\mathchar"20D$}}}}
\else{\ooalign{\hfil/\hfil\crcr\mathhexbox20D}}%
\fi}}

\begin{document}

\selectlanguage{english}

\title{Average Star Formation Parameters in the Local Volume of the Universe}
 \author{\firstname{I.~D.}~\surname{Karachentsev}}
  \email{idkarach@gmail.com}
  \affiliation\saoname

 \author{\firstname{A.~A.}~\surname{Popova}}
  \affiliation{Peter the Great St.~Petersburg Polytechnic University, Saint Petersburg, 195251 Russia}

  \begin{abstract}
  Based on the fluxes of 1400 nearby galaxies observed in far ultraviolet  (${\rm FUV}$)
   and in the H$\alpha$ line, we determined the global star formation rate per unit Universe volume,
   \mbox{$j_{\rm SFR}=(1.34\pm0.16)~10^{-2}\,M_{\odot}$\,yr$^{-1}$~Mpc$^{-3}$}. With the current star
    formation rate (${SFR}$), ($65\pm4)$\% of the observed stellar mass is reproduced in
    the cosmological time of 13.8~billion years. The neutral gas reserves in the Local
    Volume with a radius of 11~Mpc will facilitate the current ${SFR}$
  on a scale of approximately another 5~billion years.
 \end{abstract}

 \maketitle

\section{INTRODUCTION}
The integral rate of star formation in the galaxy, ${SFR}$,
expressed in units of solar masses per year
($M_{\odot}$\,yr$^{-1}$), is one of the most important
characteristics of gas transformation into stars. The specific
star formation rate per unit stellar mass, \mbox{$sSFR=SFR/M_*$}, differs
for active and passive galaxies by a factor of
\mbox{$10^4$--$10^6$}. The mean star formation rate density in a
cubic megaparsec, $j_{\rm SFR}$, determines the intensity of the
cosmic evolution of the Universe. According to multiple
observational data (Madau et al., 1998; Madau and Dickinson,
2014), the value of $j_{\rm SFR}$ in the consequential volume of
the Universe increases with decreasing redshift $z$ from $z\simeq
6$ to $z\simeq 2$ and then drops by an order of magnitude toward
the current epoch, $z=0$. In order to model the star formation
process and describe the $j_{\rm SFR}(z)$  dependences, one needs
to fix the zero point value, $j_{\rm SFR}(0)$, as accurately as
possible. This task is the main aim of our work.

\renewcommand{\baselinestretch}{0.9}
 \renewcommand{\tabcolsep}{5pt}
\begin{table*}[] \vspace*{-4mm} 
\caption{Estimates of the mean star formation rate density in the current epoch} \medskip
\begin{tabular}{c|c|r|r|r|l}
\hline
$j_{\rm SFR}(0)\times10^{-2}$   &     Data          &   \multicolumn{1}{c|}{$N$}    &    \multicolumn{1}{c|}{$z$}     &    \multicolumn{1}{c|}{$\Delta$} &   \multicolumn{1}{c}{ Reference}      \\
\hline
\multicolumn{1}{c|}{(1)}&(2)&\multicolumn{1}{c|}{(3)}&\multicolumn{1}{c|}{(4)}&\multicolumn{1}{c|}{(5)}&\multicolumn{1}{c}{(6)}\\
\hline
1.3\,$\pm$\,0.2   & H$\alpha$, $EW>10$\,\AA&   264        &  $<0.045$ &   $-$0.04~&~Gallego et al. (1995)   \\
2.2\,$\pm$\,0.7   & H$\alpha$              &   191        &   0.026   &   $-$0.03&~{P{\'e}rez-Gonz{\'a}lez}     \\
1.5\,$\pm$\,0.2   & H$\alpha$, SDSS        &$\sim100\,000$&   0.11    &   $-$0.12&~Brinchmann et al. (2004)     \\
1.1\,$\pm$\,0.3   & ${\rm FUV}$            &   896        &  $<0.10$  &   $-$0.06&~Wyder et al. (2005)     \\
 1.5\,$\pm$\,0.3  & H$\alpha$, HIPASS      &   468        &  $<0.04$  &   $-$0.03&~Hanish et al. (2006)     \\
1.6\,$\pm$\,0.2   & ${\rm FUV}$, SDSS      &$\sim50\,000$ & $\sim0.10$&   $-$0.06&~Salim et al. (2007)      \\
1.9\,$\pm$\,0.2   & H$\alpha$, SDSS        &   327        &   0.01    &   $-$0.01&~James et al. (2008)    \\
1.9\,$\pm$\,0.3   & ${\rm FUV}$, SDSS      &$\sim50\,000$ &  $<0.10$  &   $-$0.06&~Robotham and Driver (2011)  \\
1.4\,$\pm$\,0.3   & ${\rm FUV}$, H$\alpha$ &   869        &   0.001   &    0.00  &~Karachentsev et al. (2013)       \\
1.34\,$\pm$\,0.16 & ${\rm FUV}$, H$\alpha$ &   1428       &   0.001   &    0.00  &~This work    \\
\hline
\end{tabular}
\end{table*}
\renewcommand{\baselinestretch}{1.0}

To determine the $ SFR$ of a galaxy, the H$\alpha$ Balmer line
integral flux is usually used. According to Kennicutt (1998) and
Lee et al. (2009b),
$$\log SFR =8.98+2\log D + \log F_c\rm (H\alpha),$$
where the distance $D$ to a galaxy is expressed in Mpc, and the
 H$\alpha$ line flux is in erg\,cm$^{-2}$\,s$^{-1}$ and corrected
  with account for extinction in the Milky Way and in the galaxy
  itself. The details of taking into account the internal and external
  extinction are described in  Lee et al. (2009b). Another method
of estimating $SFR$ is based on measuring the flux from a galaxy
in the far ultraviolet $({\rm FUV}$, \mbox{$\lambda_{\rm
eff}=1539$\,\AA}, $FWHM=269$\,\AA) using the relation
$$\log SFR =2.78+2\log D - 0.4 m^c_{\rm FUV},$$ where $D$ is the
distance in Mpc and $m^c_{\rm FUV}$ is the apparent $\rm FUV$-band
magnitude of the galaxy corrected for internal and external
extinction  (Lee et al., 2011). The ultraviolet sky survey carried
out with the GALEX satellite serves as the main source of the
$m_{\rm FUV}$ magnitude data (Martin et al., 2005; Gil de Paz et
al., 2007).

In addition to these two methods,  $SFR$ estimates were made from the infrared flux of a galaxy in the assumption that the ${\rm FUV}$-flux of the stellar population is absorbed by interstellar dust and re-emitted in the infrared range.

\section{ESTIMATES OF THE STAR FORMATION RATE DENSITY IN THE CURRENT EPOCH}

Attempts to determine $j_{\rm SFR}(0)$ from the data on nearby galaxies were undertaken
by many authors.  Gallego et al. (1995) used for this purpose the \mbox{H$\alpha$-flux} values for 264 galaxies with  H$\alpha$ line equivalent widths larger than 10\,\AA\ and redshifts $z<0.045$. Independent H$\alpha$-flux estimates for several hundred nearby galaxies were made by {P{\'e}rez-Gonz{\'a}lez} et al. (2003)
and James et al. (2008). Hanish et al. (2006) used the equivalent widths $EW(\rm H\alpha)$ to determine $j_{\rm SFR}$ for a sample of 468~galaxies with \mbox{$z<0.04$} observed in the H\,I neutral hydrogen line in the HIPASS survey. Brinchmann et al. (2004) presented the  $j_{\rm SFR}$ estimated based on \mbox{$N\sim 10^5$} galaxies with an average redshift of $\langle z\rangle=0.11$ from the SDSS optical sky survey  (Abazajian et al., 2009).

Wyder et al. (2005) determined the local $SFR$ density from ${\rm FUV}$ fluxes of 896 galaxies with \mbox{$z<0.10$}. Salim et al. (2007) and Robotham and
Driver (2011) used for this purpose the ${\rm FUV}$-fluxes of a large number of galaxies from the SDSS survey with redshifts of $z<0.10$.

Such estimations were usually accompanied by certain expectations about the shape of the luminosity function of the galaxy sample considered and other model assumptions. An uncertainty about the size of the volume to which the measured ${\rm FUV}$-fluxes pertain has arisen during the $j_{\rm SFR}$ density calculation. The method used by  Karachentsev et al. (2013) for estimating $j_{\rm SFR}(0)$ from the Local Volume galaxies with distances of $D<11$~Mpc suffered the least from this limitation. A summary of the local star formation rate density from various sources is presented in Table~1.

Its columns contain: (1)---star formation rate density in units of
$10^{-2}\,M_{\odot}$\,yr$^{-1}$\,Mpc$^{-3}$ corrected to \mbox{$z\!=\!0$} for
the Hubble parameter \mbox{$H_0\!=\!70$\,km\,s$^{-1}$\,Mpc$^{-1}$} with
the standard error indicated;
 (2)--- the nature of the used data; (3)---the number of galaxies
in the sample; (4)---the redshift mean value and interval; (5)---
$z=0$ correction in the density logarithm $\Delta\log(j_{\rm
SFR})$ in an assumption that the
 \mbox{$j_{\rm SFR}(z)/j_{\rm SFR}(0)\!\simeq\!(1\!+\!z)^{2.7}$}
dependence is true for \mbox{$z\ll1$} according to Madau
and Dickinson (2014); (6)---reference to the observational data source.

The data presented above demonstrates that the difference between
some average star formation rate density estimates in the current
epoch reaches a factor of~2.

\section{SAMPLE OF LOCAL VOLUME (LV) GALAXIES}
The list of nearby galaxies with expected distances not exceeding
11~Mpc has been replenishing quickly in the recent years due to
the emergence of deeper optical and H\,I-surveys of large areas of
the sky. While the Catalog of Neighboring Galaxies (Karachentsev
et al., 2004) included \mbox{$N=450$} galaxies, in the Updated
Nearby Galaxy Catalog (UNGC, Karachentsev et al., 2013), their
number has increased to \mbox{$N=869$}. Observational data on the Local
Volume galaxies are presented online at {\url
{http://www.sao.ru/lv/lvgdb}} and described in Kaisina et al.
(2012). Distances were measured for more than half of the sample
galaxies with the Hubble Space Telescope, at an accuracy of about
5\%. The most complete summary of the distances is presented in
the Extragalactic Distance Database = EDD (Anand et al., 2021).
The online version of the UNGC catalog is regularly updated by new
objects and new data on the nearby galaxies. Currently, the number
of suggested Local Volume members exceeds 1400~objects.

\begin{figure*}[] \vspace*{-4mm}
\includegraphics[scale=0.9]{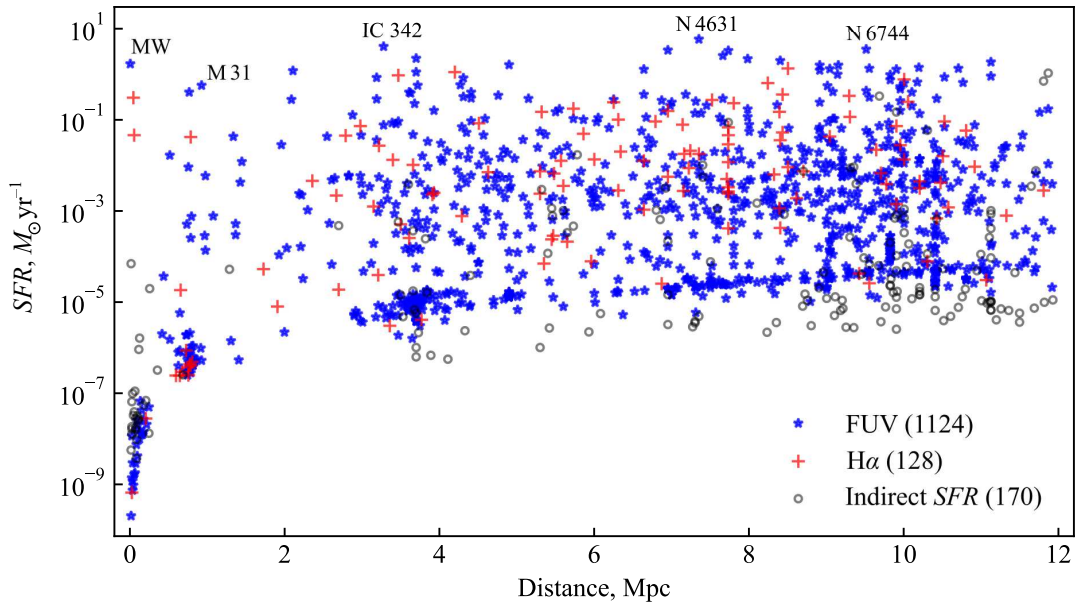}
 \caption{Distribution of the Local Volume galaxies by integral star formation rate $SFR$ and distance. Galaxies with $SFR$ estimates based on the ${\rm FUV}$-flux are shown by stars, those with estimates from the H$\alpha$-flux are shown by crosses, and objects with indirect $SFR$ estimates by the calibration dependence are marked by empty circles.}
\end{figure*}

A significant portion of these galaxies have the H$\alpha$- and
${\rm FUV}$-fluxes measured. When determining the integral star
formation rate of a galaxy, we preferred ${\rm FUV}$-flux data,
since the $SFR$ estimates based on these data relate to a typical
time interval of about 100~million years, whereas the H$\alpha$-flux
of a galaxy characterizes the $SFR$
 on a shorter scale of about 10~million years. We used the $\rm
SFR(H\alpha)$ estimates for galaxies with no \mbox{${\rm
FUV}$-flux} data. Usually this applied to objects at low galactic
latitudes, where the \mbox{${\rm FUV}$-range} extinction is especially
high.

Note that the ${{SFR\,(\rm H\alpha)}}$ and ${{SFR\,(\rm FUV)}}$
estimates are mutually well-calibrated for the case %\linebreak 
of spiral
galaxies. When transitioning from massive galaxies to dwarfs, the
dispersion of the \linebreak \mbox{${{SFR({\rm H\alpha})/SFR \rm (FUV)}}$} ratio
increases with decreasing stellar mass of a galaxy, and the
average value of this ratio plummets approximately by a factor of
three for the least massive objects (Lee et al., 2009a;
Karachentsev and Kaisina, 2019).

Along with the $SFR$ values for the Local Volume galaxies, the
online version of the UNGC contains data on the stellar masses
$M_*$ and masses of neutral hydrogen $M_{\rm HI}$ in these
galaxies. Both these parameters are the main integral markers of
the gas-to-stars transformation process.

\section{STAR FORMATION RATE IN THE LOCAL VOLUME}
The distribution of LV galaxies by integral star formation rate and distance within 12~Mpc is presented in Fig.~1.
Galaxies with $SFR{\rm (FUV)}$
estimates are shown by stars, and objects with $SFR$
measured only by H$\alpha$-flux are shown by crosses. ${\rm FUV}$ and H$\alpha$-flux data are missing for 170 sample galaxies. The $SFR$
estimates for these galaxies are made by calibration dependences between $SFR$
and $M_*$, constructed for galaxies of various morphological types (presented below in Section~5). These galaxies are shown in Fig.~1 by the empty circles.

About a quarter of the LV galaxies have only the upper ${\rm FUV}$-flux estimates available. Most of them are dwarf spheroidal galaxies (dSph) with no noticeable gas stored for star formation. We attributed to them the $SFR$ value
corresponding to the upper limit of the ${\rm FUV}$-flux. These faint objects form a narrow sequence in the lower part of Fig.~1 near \mbox{$SFR= 10^{-5}$}. Taking these galaxies into account or ignoring them changes the average $SFR$
density estimate in the LV by only a fraction of a percent.

For the most extended galaxies, the Milky Way (MW) and Andromeda (M\,31), ${\rm FUV}$-fluxes were taken from Chomiuk and Povich (2011) and Rahmani
et al. (2016) correspondingly. Evidently, the LV hosts 10~galaxies with star formation rates higher than in our Galaxy.

The brightest star formation source is in the spiral galaxy NGC\,4631, which reproduces about $5\,M_{\odot}$ per year. At $D>10$~Mpc distances the $SFR$ data density
drops rapidly, which points to the incompleteness of the sample at the far LV boundary.

\begin{figure*}  \vspace*{-6mm}
\includegraphics[scale=0.9]{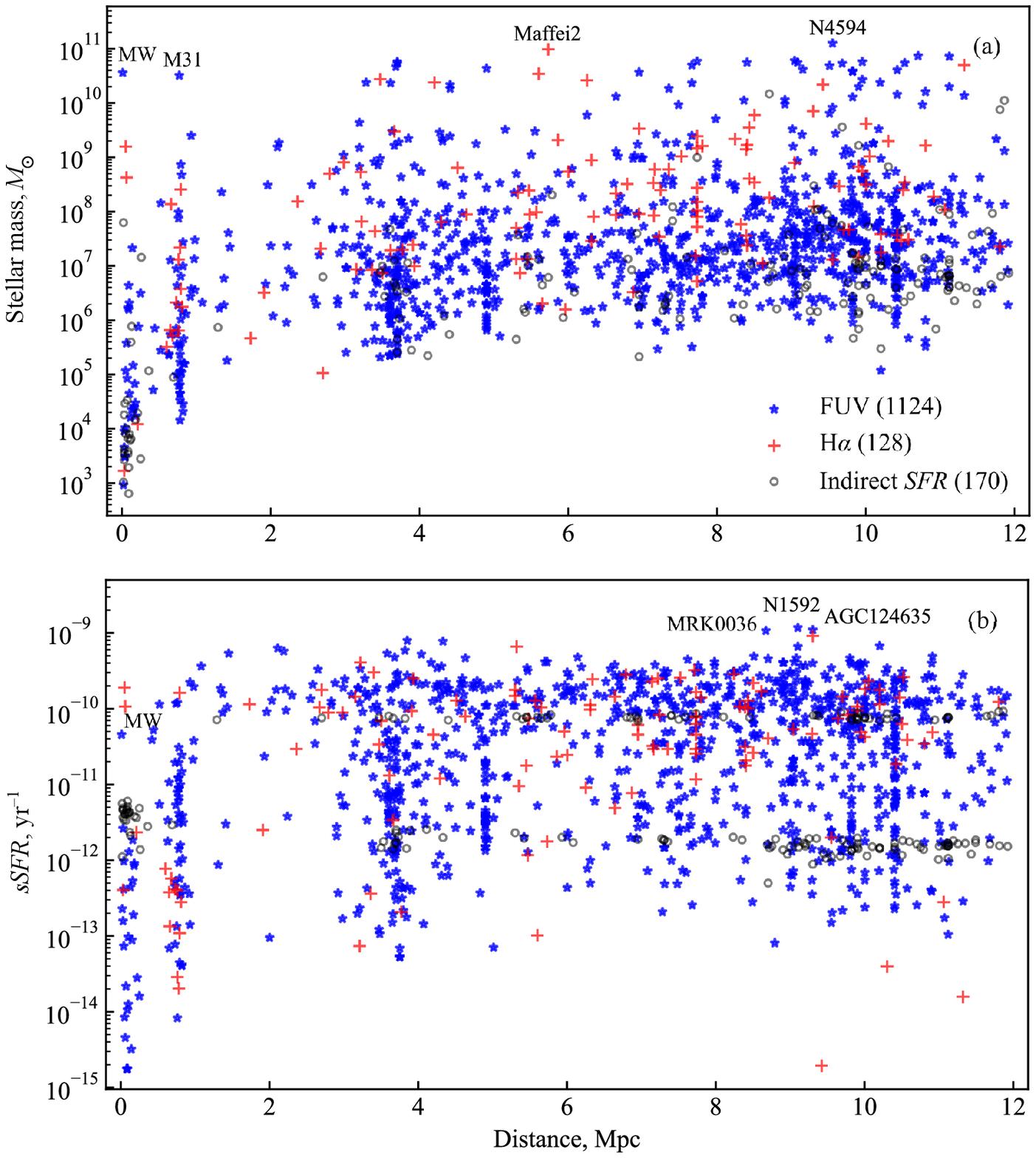}
\caption{
Panel~(a)---distribution of the Local Volume galaxies by stellar mass and distance.
Panel~(b)---distribution of the same galaxies by specific star formation rate. The symbols are the same as in Fig.~1.}
\end{figure*}

\begin{figure*} \vspace*{-6mm}
\includegraphics[scale=0.9]{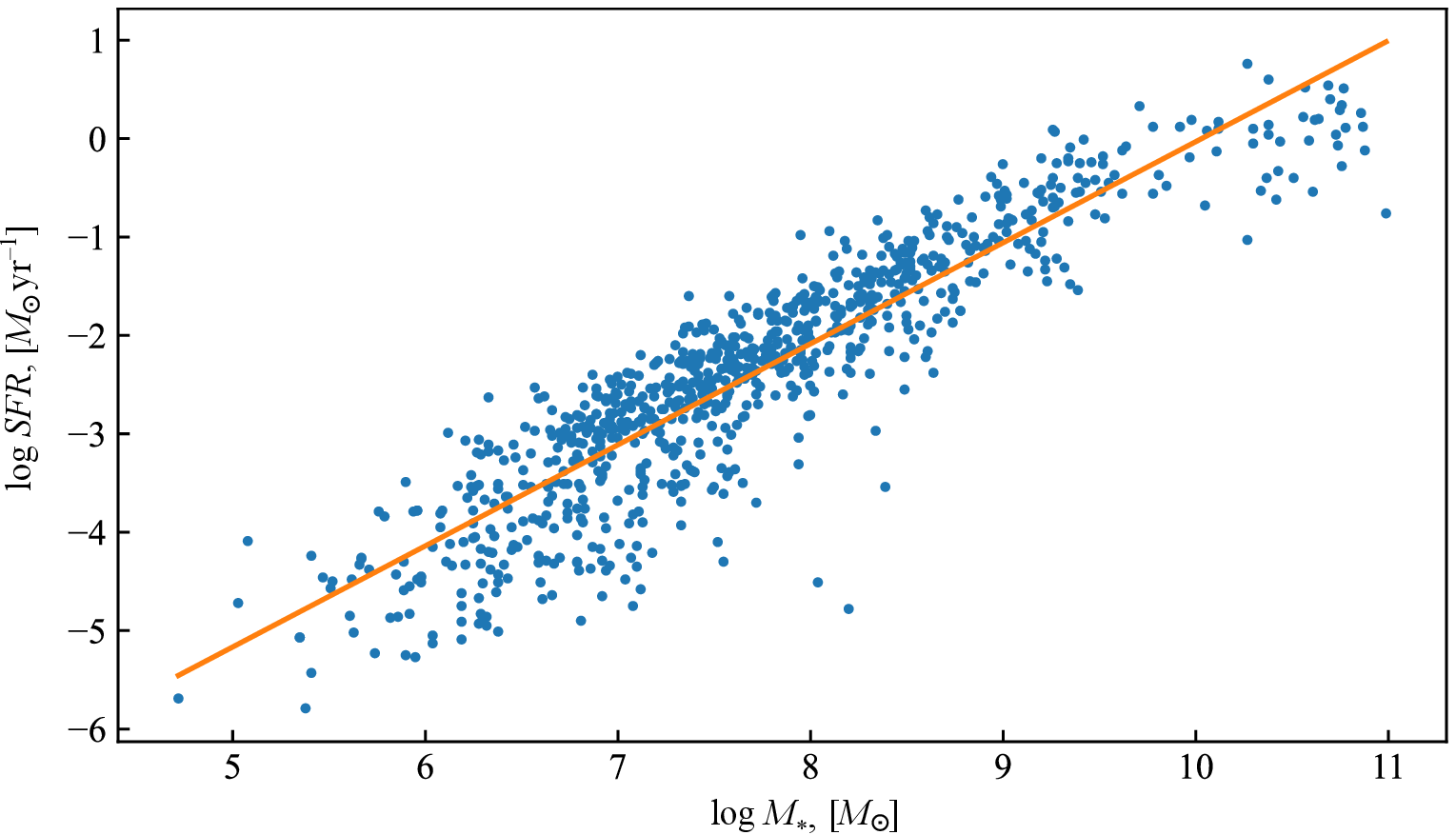}
\caption{Relation between the integral star formation rate and stellar mass for late-type Local Volume galaxies ($T=2$--$10$).}
 \end{figure*}

\section{SPECIFIC STAR FORMATION RATE IN LV GALAXIES}
Determining the specific star formation rate of a galaxy $sSFR= SFR/M_*$
presumes the knowledge of its total stellar mass. However, thorough
integral mass estimates are known for only a few nearby galaxies.
 Calculating them requires using different model assumptions, in particular,
 about the form of the initial stellar mass function. A simplified method of estimating $M*$ is based on the data on integral galaxy luminosity, preferably in the $K$-band, where the effects of internal light extinction and bursts of star formation are minimal.

The UNGC catalog  (Karachentsev et al., 2013) contains data on the luminosity of galaxies in the \mbox{$K$-band}, $L_K$, measured in the 2MASS-survey of the entire sky  (Jarrett et al., 2000; Skrutskie et al.,
2006) or estimated from the luminosity of a galaxy in other bands. To determine $M_*$ we used the relation
\mbox{$M_*/L_K=0.6\,M_{\odot}/L_{\odot}$}, justified by Lelli
et al. (2016). The proportionality coefficient 0.6 depends weakly on the morphological type of a galaxy and was neglected.

The distribution of LV galaxies by stellar mass and distance is presented in Fig.~2a.
 The same symbols as in Fig.~1 are used. Evidently, the LV includes 22 galaxies with masses exceeding the stellar mass of the Milky Way and M\,31. The most massive representatives of the LV are NGC\,4594 (``Sombrero'') and Maffei\,2, relating to the early types with suppressed star formation.

Figure~2b shows the distribution of LV galaxies by specific star formation rate and distance. The designations of galaxies with different $SFR$ data sources are the same as in the upper panel.

The highest star formation rates per unit stellar mass are
demonstrated by the dwarf galaxies (NGC\,1592, AGC\,124635,
Mrk\,036) in an \linebreak active phase (burst) of star formation. The
diagram shows horizontal sequences of ``active'' galaxies
\mbox{($sSFR\sim10^{-10}$~yr$^{-1})$}, ``passive'' galaxies
 \mbox{($sSFR\sim10^{-12}$~yr$^{-1})$} with the so-called ``green valley''
between them.

The vertical ``columns'' in Fig.~2 correspond to groups of
galaxies where the distances of some faint members are identified
with the distance to the main galaxy.

Figure~3 shows the distribution of spiral and irregular galaxies
with morphological types \mbox{$T =2$--$10$} on the de Vaucouleurs scale
by integral star formation rate and integral stellar mass. The
linear regression there, $\log{SFR} = 1.02 \log {M_*} - 10.30$,
was used to estimate the $SFR$
in galaxies without ${\rm FUV}$- and \mbox{H$\alpha$-fluxes},
which are shown by the empty circles in the previous figures. A
similar calibration dependence,
${\log SFR = 0.86\log(M_*) - 10.80}$,
was constructed for passive galaxies \mbox{$(T<2)$} in order
to account for their very small contribution to the total $SFR$
of the Local Volume.

\begin{figure*}  \vspace*{-5mm}
\includegraphics[scale=0.88]{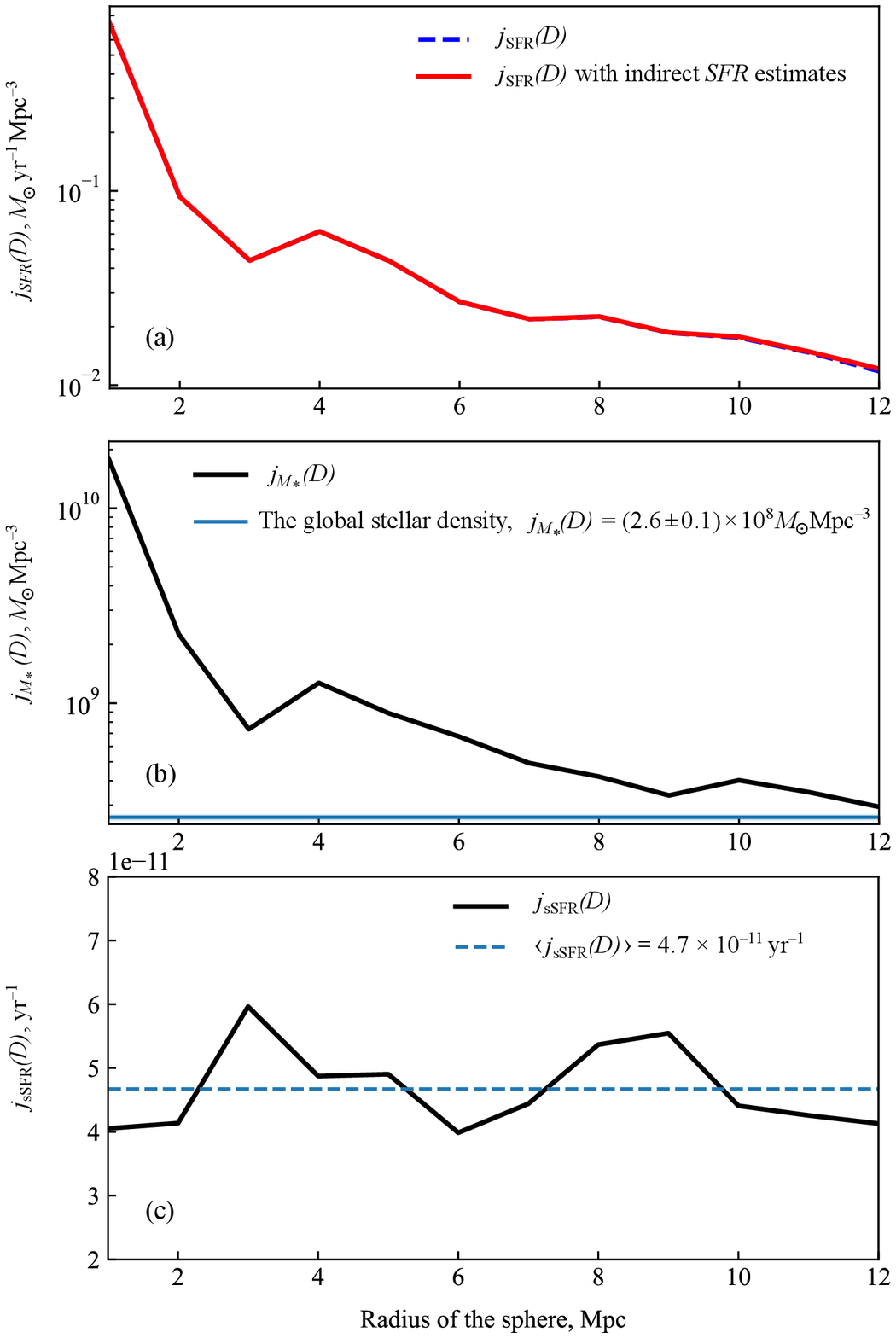}
\caption{ Panel~(a)---mean star formation density in the Local Volume
spheres of different radii~$D$. Panel~(b)---mean stellar mass
density in the Local Volume spheres. The horizontal stripe near
the lower edge corresponds to the global stellar density.
Panel~(c)---distribution of the  specific star formation rate
in the Local Volume.}
\end{figure*}

\section{SFR DENSITY IN THE LOCAL VOLUME}
Summing up the $SFR$ values for individual galaxies and dividing the sum by volume ($4\pi/3) D^3$, we obtain
the distribution of the mean star formation rate density within the distance $D$, shown in
Fig.~4a. The density $j_{\rm SFR}(D)$ drops with increasing $D$ for two reasons. First, we
are located inside the Local Group, which is surrounded by several other groups (M\,81, NGC\,253, Centaurus\,A)
 at a distance of about 4~Mpc. Second, the completeness of the sample decreases near
  the sample boundary at \mbox{$D\gtrsim10$}~Mpc due to the missed galaxies whose
  distances remain unmeasured. The upper polygonal line shows the contribution of all
  LV galaxies with account for the 170~galaxies with no measured ${\rm FUV}$- and
  H$\alpha$-fluxes. The contribution of these mostly small galaxies is noticeable only at the far LV boundary.

Figure~4b shows the distribution of the average stellar mass
density  $j_{\rm M_*}$ in the LV. Evidently, the   $j_{\rm
SFR}(D)$ and $j_{\rm M_*}(D)$ distributions behave in a 
similar
manner. The horizontal stripe near the lower edge of the panel
corresponds to the average global stellar density
\mbox{$j_{M_*}\!=\!(2.6\!\pm\!0.1)\!\times\!10^8~M_{\odot}$\,Mpc$^{-3}$}, 
obtained
from the Driver et al. (2012) data %\linebreak 
regarding the  \mbox{$K$-luminosity}
global density,
%\linebreak 
\mbox{$(4.3\!\pm\!0.2)\!\times\!10^8\,L_{\odot}$\,Mpc$^{-3}$}, for
$M_*/L_K\!=\!0.6\,M_{\odot}/L_{\odot}$. The observed distribution of
stellar density in the LV is higher than the global value on all
scales. According to the data obtained by Karachentsev and
Telikova (2018), the excess of the observed local stellar density
in an 11~Mpc radius volume amounts to a factor of $(1.46\pm0.10)$
and becomes insignificant at $D\simeq40$~Mpc.

Figure~4c replicates the behavior of the mean specific star formation rate with increasing distance $D$. The variations of $j_{\rm sSFR}$ with increasing volume are relatively small.

\begin{figure*}  \vspace*{-5mm}
 \includegraphics[scale=0.9]{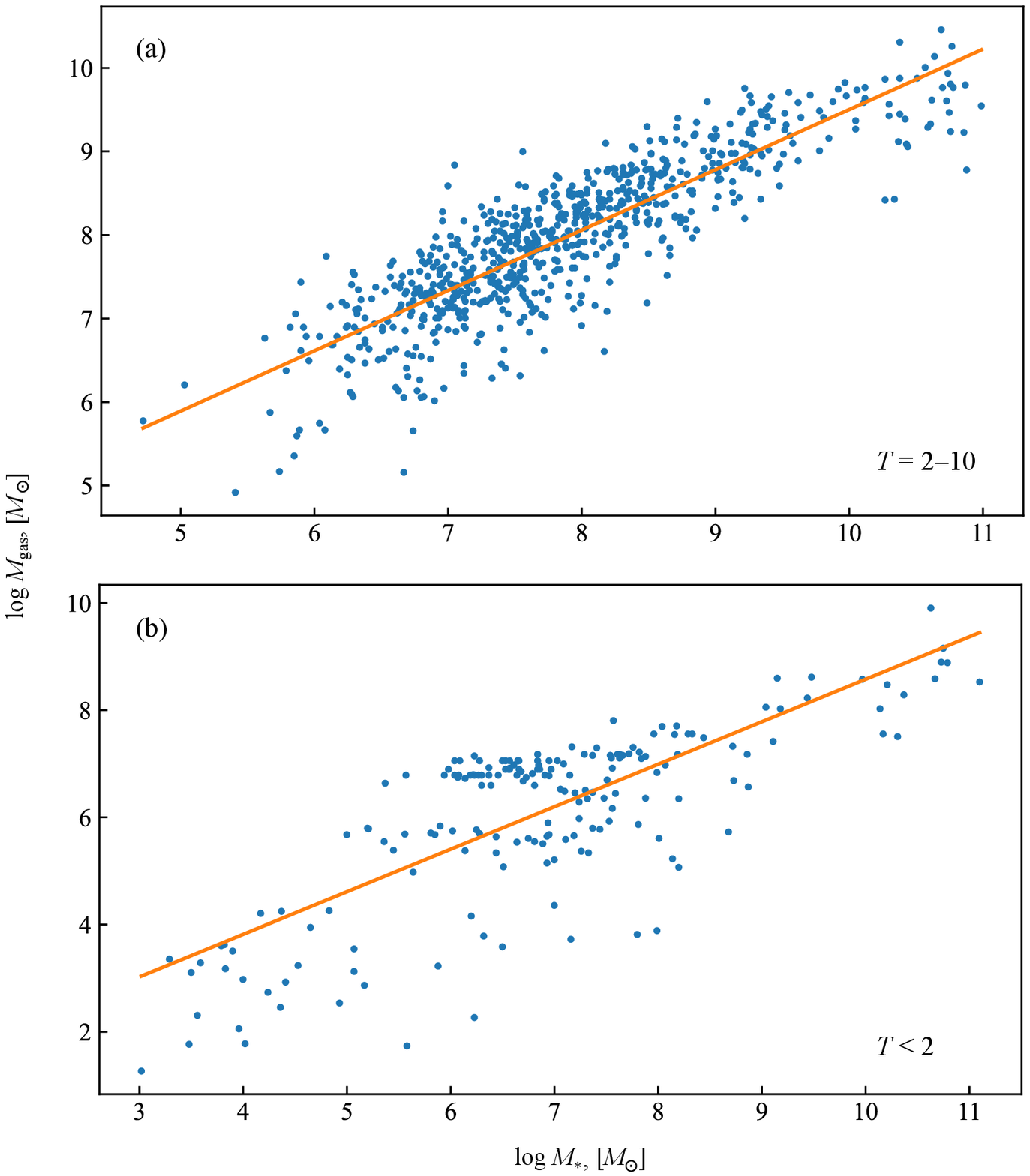}
 \caption{Relation between the gas mass and stellar mass for late-type $T=[2$--$10]$ (a) and early-type galaxies ${T<2}$ (b).}
\end{figure*}
We adopted $\langle j_{\rm sSFR}\rangle=(4.7\pm0.3)~10^{-11}$~yr$^{-1}$ as the mean value. Therefore, given this average star formation rate, the Local Volume reproduces $(65\pm4)$\% of its observed stellar mass over the cosmological time of 13.8~billion years.

\section{AVERAGE DENSITY OF NEUTRAL GAS IN THE LOCAL VOLUME}
Surveys of large areas of the sky in the neutral hydrogen H\,I line at
$\lambda=21$~cm (HIPASS, Koribalski et al., 2004; ALFALFA,
Haynes et al., 2011), as well as dedicated observations of individual galaxies in  H\,I (Huchtmeier et al., 2000) led to the discovery of an H\,I-flux, $F_{\rm HI}$, in most of the nearby galaxies.

According to Roberts and Haynes (1994),
the mass of neutral gas in a galaxy is expressed through $F_{\rm HI}$ as
$M_{\rm gas} =1.4\times2.36 \times10^5 D^2~F_{\rm HI}$,
where $D$ is measured in Mpc, $F_{\rm HI}$ is in  Jy\,km\,s$^{-1}$, and the 1.4 factor accounts for the contribution of heavier elements to the gas mass.

The panels in Fig.~5 show the relation between the gas mass and stellar mass in early type galaxies ($T<2$) and in spiral and irregular galaxies with morphological types $T=(2$--$10)$.

\begin{figure*} 
 \includegraphics[scale=0.86]{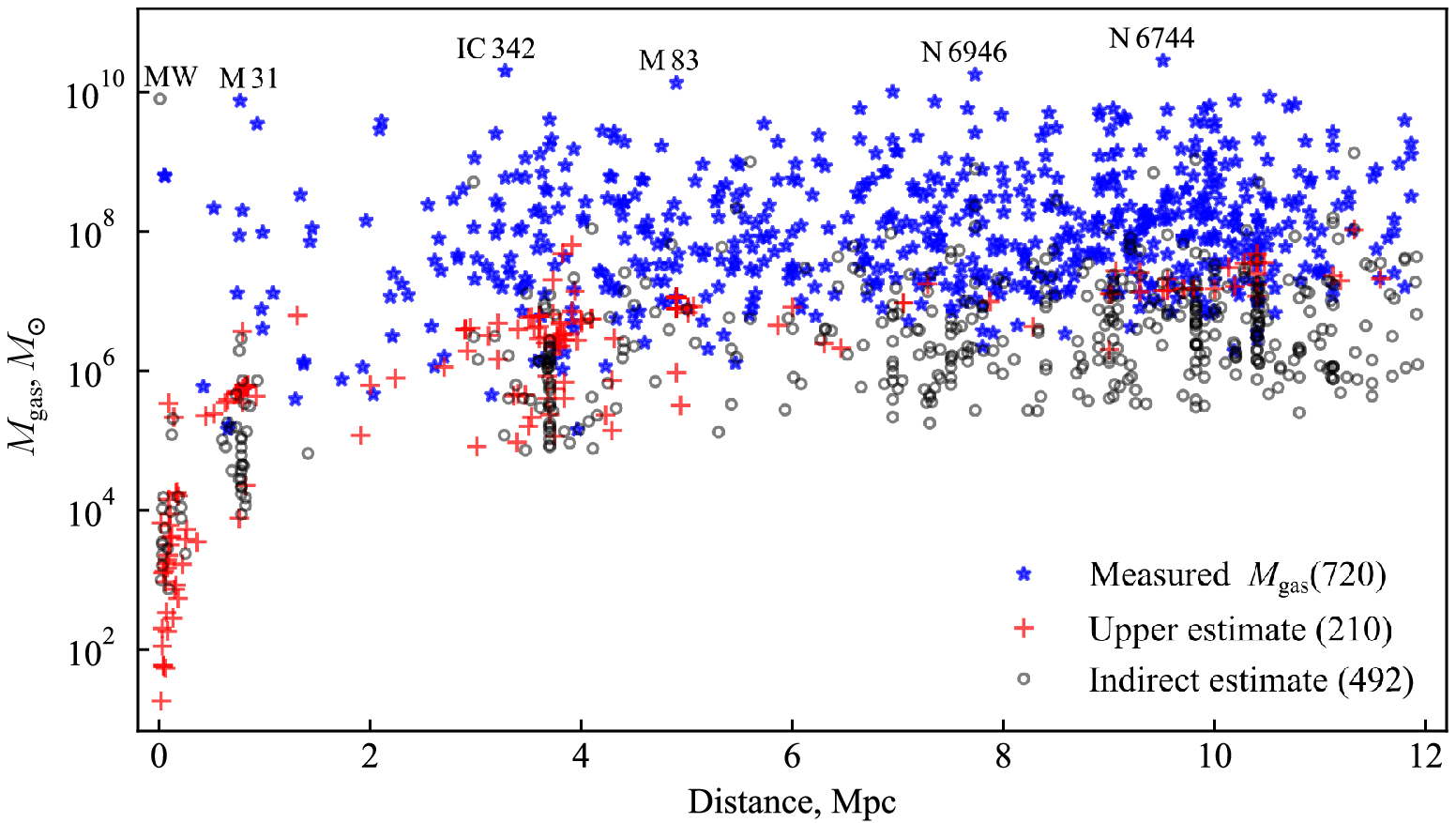}
 \caption{Distribution of the Local Volume galaxies by gas mass and distance. Galaxies with direct  $M_{\rm gas}$ estimates are designated by stars, galaxies with upper H\,I-flux limits are shown by crosses, and galaxies with indirect  $M_{\rm gas}$ estimates obtained from calibration dependences are shown by the empty circles.}
\end{figure*}

These calibration dependences with linear regressions $\log M_{\rm gas} = 0.78\,\log {M_*}+ 0.80$ for {$T<2$} and
 ${\log M_{\rm gas} = 0.72\,\log M_* + 2.30}$ for \mbox{$T=(2$--$10)$} were used to estimate the mass of the gas in the LV galaxies that were not covered by the H\,I observations. This primarily relates to northern sky galaxies with declinations Dec~$>+39^{\circ}$, located beyond the HIPASS and ALFALFA survey limits. The regression lines in both panels show that the $M_{\rm gas}/M_*$ ratio increases from massive galaxies to dwarfs, indicating a slower (dormant) star formation process in dwarf systems.

The distribution of LV galaxies by gas mass and distance is presented in Fig.~6. Only half of the LV galaxies (720 out of 1422) have direct H\,I-flux measurements. For the remaining 492~galaxies (shown by open circles) the gas mass estimates were made using the calibration dependences in Fig.~5. The objects richest in gas are the spiral Sc-galaxies NGC\,6744, IC\,342, NGC\,6946.

\begin{figure*} 
 \includegraphics[scale=0.84]{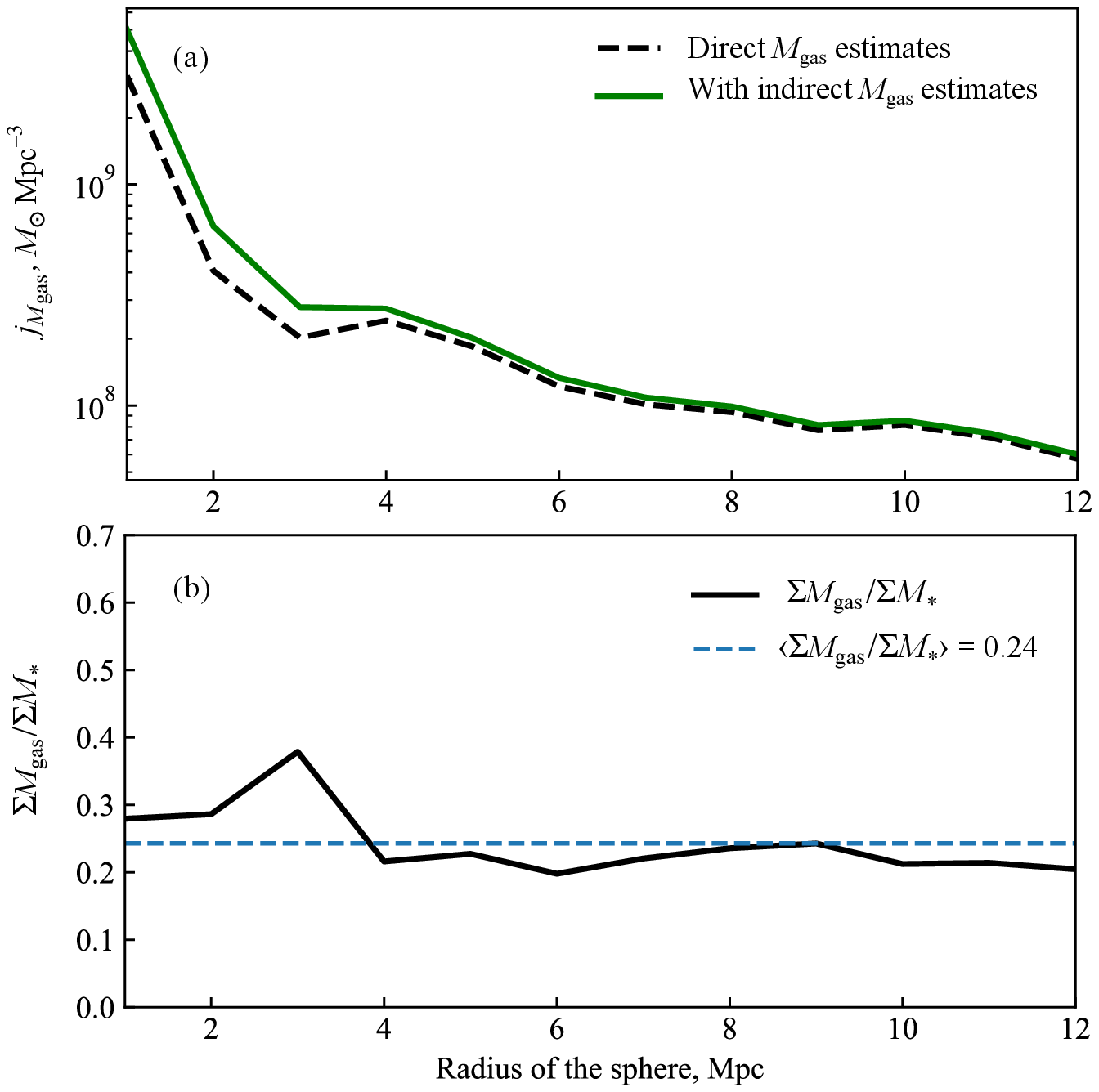}
 \caption{
 Panel~(a)---dependence of the mean gas mass density on distance in the Local Volume. The difference between the upper and lower lines corresponds to the contribution of galaxies with indirect mass estimates, %?????? ??????
 panel~(b)---average gas-to-stellar mass ratio in the Local Volume spheres of different radii.}
\end{figure*}

\begin{figure*}
\includegraphics[scale=0.83]{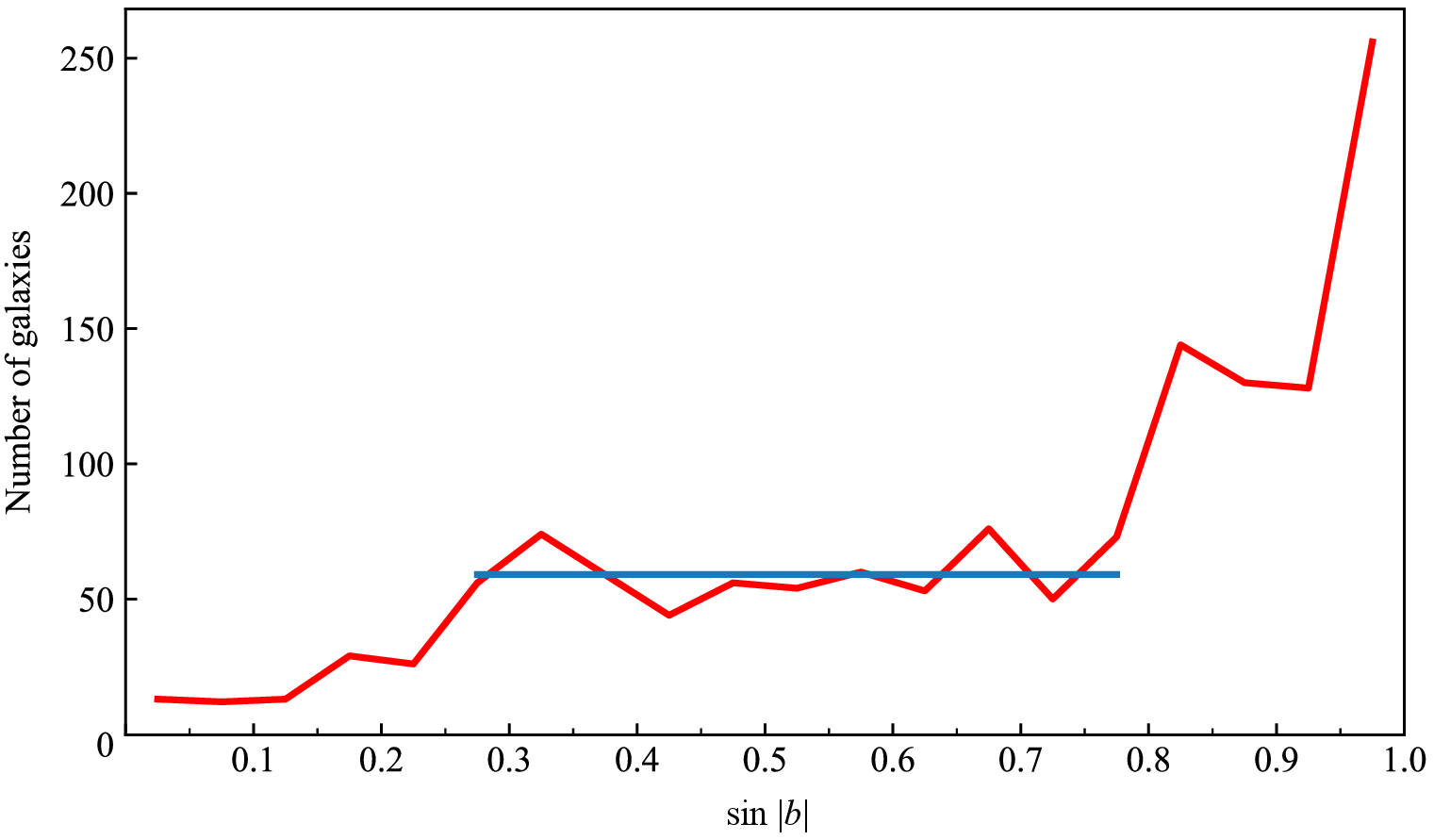}
 \caption{Distribution of the number of Local Volume galaxies in galactic latitude sine intervals with a step of 0.05.
 The horizontal line corresponds to the assumed values free from the Local supercluster effects and extinction in the Milky Way.}
\end{figure*}
Figure~7a shows the distribution of the average neutral gas mass density in  $D$~Mpc radius spheres. The lower polygonal line shows the dependence with account for only the direct  H\,I-flux estimates. The upper line is plotted for the entire galaxy sample, including cases with gas mass estimates made by the calibration dependences in Fig.~5. The contribution of these unobserved galaxies to the total gas density turned out to be small.

The total gas-to-stellar mass ratio for galaxies inside spheres of different radii is presented in Fig.~7b. The typical average gas-to-stars density ratio in the LV is approximately 0.24. Therefore, the available stock on neutral gas in the LV can maintain the observed star formation rate in it on a scale of another 5~billion years. We should note, however, that according to the Zhou et al. (2023) data, roughly half of the neutral gas in the galaxies can be below the sensitivity limit of the carried out H\,I-surveys.

The small value of the $M_{\rm gas}/M_*$ ratio indicates that the star formation process in the Local Volume is already fading, nearing its completion.

\begin{figure*} 
    \includegraphics[scale=0.81,angle=0]{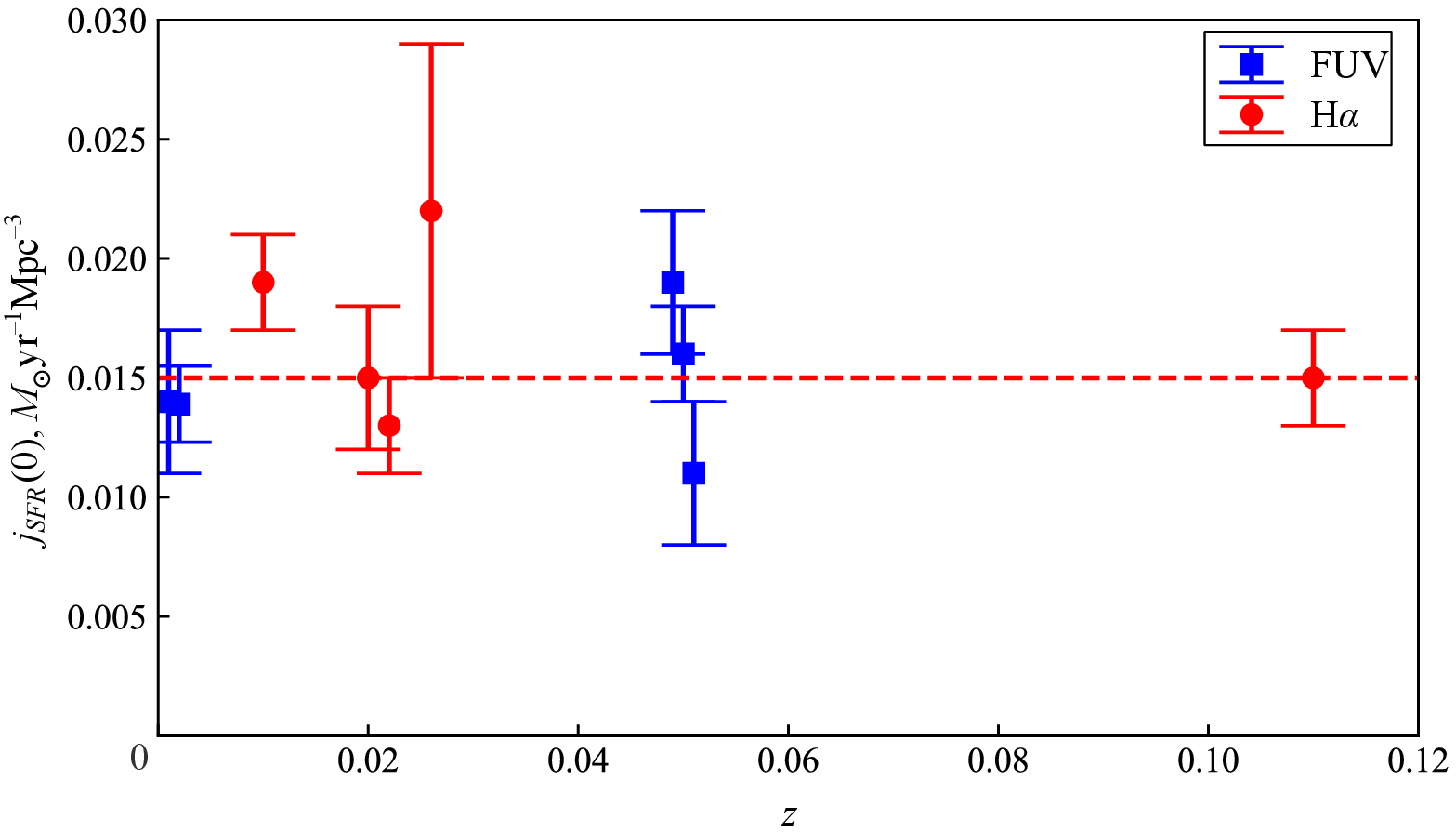}
 \caption{Estimates of global star formation rate per unit Universe
  volume based on the  ${\rm FUV}$-flux (squares) and H$\alpha$-flux (circles) data,
  reduced to the $z=0$ epoch, with the Hubble parameter $H_0=70$~km\,s$^{-1}$~Mpc$^{-1}$.
   The horizontal  dashed %dotted
   line corresponds to the average value of this quantity according to  Madau and Dickinson (2014).}
\end{figure*}

\section{DISCUSSION AND CONCLUDING REMARKS}
The data on the $SFR, M_*$  and $M_{\rm gas}$ density presented in Fig.~4 and Fig.~7 were obtained in the assumption of a homogeneous distribution of LV galaxies on the celestial sphere. At the same time, the number of galaxies in the sine intervals of galactic $\sin b$ should be the same. The real distribution of nearby galaxies is distorted by the strong light extinction in the Milky Way zone, as well as by the concentration of galaxies in the Local supercluster plane, which is almost perpendicular to the plane of the Milky Way. Figure~8 shows the distribution of the number of LV galaxies by  $\sin|b|$ intervals with a step of 0.05.

The approximate constancy of the number of galaxies in these
intervals is true only at the average galactic latitudes in
the $|b|=15^{\circ}$--$50^{\circ}$ range. At high latitudes, the
excess of the number of galaxies is due to the effect of the Local
supercluster. The expected lack of galaxies at $|b|<15^{\circ}$ is
caused by the absorption of light by interstellar dust, which is
especially strong in the ${\rm FUV}$-range. Extrapolating the
horizontal line in Fig.~8 to the regions of low galactic
latitudes, we derive the shortage in the total number of LV
galaxies \mbox{$\Delta N=258$}. Therefore, the density estimates
presented in Fig~4a, should be multiplied by a factor of
\mbox{$(1428+258)/1428=1.18$}. Taking this factor into account, we
adopt the average $SFR$ density within the distance
\mbox{$D=11$}~Mpc equal to
\mbox{$(1.95\pm0.18)\times10^{-2}~M_{\odot}$\,yr$^{-1}$\,Mpc$^{-3}$}.
Additionally, since the stellar mass excess in this volume
relative to the global value amounts to a factor of $1.46\pm0.10$
(Karachentsev and Telikova, 2018), for the global star formation
rate in unit volume we derive the estimate \mbox{$j_{\rm
SFR}=(1.34\pm0.16)~10^{-2}~M_{\odot}$\,yr$^{-1}$\,Mpc$^{-3}$}, which is
presented in the last column of Table~1.

We should note, however, that the 1.18 factor is obtained in the
assumption that the number of unaccounted for galaxies at low
galactic latitudes does not depend on their mass (luminosity).
This assumption may turn out to be disputable if the percentage of
dwarf galaxies obscured from us is significantly higher than that
of massive spiral galaxies that contribute the most to the
integral star formation rate. Evidently, the effect of this
circumstance on the final result is still small.

The mean star formation rate density in the Universe from Table~1
corrected for epoch $z=0$ is replicated in Fig.~9 with the mean
errors indicated. Estimates made from H$\alpha$- and ${\rm
FUV}$-fluxes are shown by, correspondingly, circles and squares.
The horizontal dotted line indicates the average value \mbox{$j_{\rm
SFR}(0)=0.015\,M_{\odot}$yr$^{-1}$\,Mpc$^{-3}$}, fixed by  Madau and Dickinson
(2014)  as the most optimal.

Evidently, the $j_{\rm SFR}(0)$ estimates for %\linebreak
\mbox{$H_0=70$~km\,s$^{-1}$~Mpc$^{-1}$} and with a correction for
the effect of evolution are in good agreement with each other
within the margin of measurement error. Systematic differences
between the $j_0(SFR)$ values obtained from H$\alpha$- and ${\rm
FUV}$-fluxes are small. The estimates of this value made from
infrared galaxy fluxes (Sanders et al., 2003; Takeuchi et al.,
2003) also agree with the optical data within errors. We can
assume that at present, the estimates of the cosmic star formation
density at $z=0$ coincide with an accuracy of about 10\%.

To conclude, let us note that the analysis and modeling of  $SFR$
and $M_*$ observational data for the Local Volume galaxies, performed
recently by Haslbauer et al. (2023), change noticeably the history of
cosmic star formation in the Universe, making the $SFR(z)$ dependence
more flat and lowering the peak in the Madau diagram for $z\sim2$
by a factor of 2.2.
 Haslbauer et al.
(2023) used the UNGC catalog data on the average  $SFR$ and $M_*$
density to test various star formation history models in massive and
dwarf galaxies. The authors have shown that the star formation rate
evolution model in unit volume, $ {j_{\rm SFR}(z)}$, with a steep
peak at ${z\sim2}$, presented by  Madau et al. (1998) and
Madau and Dickinson (2014), contradicts the observed average stellar
mass density in the LV. In order to reconcile the star formation
history in the nearby galaxies with the observed amount of their
stellar mass, one must assume that the real burst of star formation in the $ {z\sim2}$ epoch was 2--3~times less intensive than that predicted by the classic  Madau diagram.

%\section*{CONFLICT OF INTEREST}
%The authors of this work declare no conflicts of interest.

\begin{acknowledgments}
This work has made use of the Local Volume galaxy database, which is updated with the support of grant No.~075-15-2022-262 (13.MNPMU.21.0003) of the Ministry of Science and Higher Education of the Russian Federation.
\end{acknowledgments}

\end{document}